# Complementary Normalized Compressive Ghost Imaging with Entangled Photons

Dawei Liu,[1, 2] Lifei Li,[1, a]

[1]*State Key Laboratory of Transient Optics and Photonics, Xi'an Institute of Optics and Precision Mechanics, Chinese Academy of Sciences, Xi'an 710119, China*

[2]*University of Chinese Academy of Sciences, Beijing 100049, China*

[a]*Authors to whom correspondence should be addressed. Electronic addresses: lilifei@opt.ac.cn*

We demonstrate a compressive normalized ghost imaging system with entangled photons employing complementary compressive imaging (CCI) technique. The quantum ghost image reconstruction was achieved at only 19.53% sampling ratio of raster scanning. With the special +1/-1 type sensing matrix and appropriate optimal algorithm, the photon utilization efficiency and robustness of the imaging system is enhanced significantly. Our results reveal the great potential of CCI technique applied in quantum imaging and other quantum optics field such as quantum charactering and quantum state tomography to use the information loaded in each photon with high efficiency.

Ghost imaging (GI)[1], also called correlated imaging, has attracted much attention for its huge applications in remote sensing[2,3], optical encryption[4,5] and secure image transmission[6]. In general, this intriguing nonlocal imaging protocol allows one to employ two spatial correlation beams to reproduce the image of an object in a path that never directly interacts with the object. The light sources of GI can be mainly itemized into two categories: one is classical thermal light (i.e., T-type GI)[7-9], and another one is entangled-photons (i.e., E-type GI) generated by spontaneous parameter down conversion (SPDC)[10-13]. In theory, the entangled photons could offer higher imaging quality in contrast to its thermal light counterpart in GI[14]. However, the E-type GI typically requires long sampling time when it comes to scanning point by point due to the ultra-low photon flux. Compressive sensing (CS) is an advanced signal sampling and reconstruction theory that beats the limits of Nyquist-Shannon criterion using natural (or under proper representation basis) sparsity of the original signal[15,16]. It allows one to reconstruct the original signal from a series of random projective measurements by optimal algorithm, resulting in the sampling ratio is greatly decreased. Combining CS theory with the correlation property of entangled photons can significantly increase the photon utilization efficiency and strengthen the robustness of imaging system, which we believe is a vital step toward diversified application of E-type GI. However, thus far, few papers have been reported on compressive E-type GI: In 2011, Zerom *et al.* demonstrated the operation of compressive E-type GI with a spatial light modulator (SLM)[11]. But the SLM is a typical phase-only modulator that makes the imaging system complicated for the reason of mimicking amplitude-only modulation. Later in 2013, Loaiza *et al.* reported a quantum compressive object tracking[12], in which a digital micro-mirror device (DMD) was used. However, at each measurement, only about half of photons falling on DMD are collected in their system, which reduces the photon utilization efficiency. Therefore, it is crucial to explore and develop novel E-type GI with simpler configuration and better performance (e.g., faster imaging time, higher photons utilization efficiency and lower sampling ratio).

Recently, complementary compressive imaging (CCI) technique for classical information processing field has been demonstrated as a promising sampling approach with superior performance[17-19]. By making full use of the double reflection direction of DMD or loading complementary positive-negative random patterns onto DMD, the +1/-1 type sensing matrix is ingeniously constructed and can better satisfy the restricted isometry property (RIP)[20]. The quality of reconstructed image with CCI technique is greatly improved compared with the traditional compressive imaging based on 0/1 type sensing matrix. Inspired by the work 11-12 and 17-19, we firstly transplant the CCI into the quantum imaging context and experimentally investigate its effect.

In this letter, we report a CCI-based E-type GI system by using a DMD and two bucket detectors (without spatial resolution). This system reduces both the measurement numbers and photons required to form a quantum ghost image, thereby being faster and more efficient than common E-type GI system. In addition, we also investigate the quality of quantum ghost image reconstructed by two different reconstruction algorithms: orthogonal matching pursuit (OMP)[21] and total variation minimization by augmented Lagrangian and alternating direction (TVAL3)[22]. The experimental results show that TVAL3 algorithm is more sensitive to RIP, and has a better performance than OMP algorithm. This opens the possibility of CCI technique using in other quantum optics filed such as quantum charactering[23] and quantum state tomography[24].

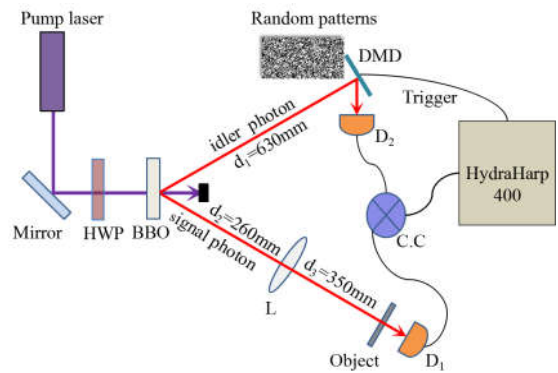

Fig. 1. Experiment setup for complementary normalized compressive quantum ghost image: the HWP, half wave plate;

BBO, β-barium borate crystal; L, the imaging lens; DMD, digital micro-mirror device; the Random patterns represent the complementary two-dimensional random binary patterns impressed on DMD; $D_1$ and $D_2$ are bucket detectors; C.C represents coincidence measurement between $D_1$ and $D_2$; $d_1$, $d_2$ and $d_3$ is the distance between BBO and DMD, BBO and L, L and Object, respectively.

Our quantum imaging system is sketched in Fig.1. A 405 nm, 50 mw, continuous-wave laser vertically pumps a type II phase matched 2 mm-thickness BBO crystal. The half wave plate (HWP) in front of the BBO crystal is used to adjust the polarization of pump laser to maximizing the photon down conversion efficiency. The degenerated polarization-position-entangled photon pairs with 810 nm center wavelength are produced by SPDC at the 2.66 degree direction respected to pump laser. One of the photon pair, the signal photon is collected by bucket detector $D_1$ after passing through imaging lens $L$ ($f$=250 mm) and object (a 200 μm width double slit with center-to-center separation 800 μm). Another one, the idler photon impinges on the DMD, then is reflected and collected by bucket detector $D_2$. The $D_1$ and $D_2$ are consisting of a collection lens with 25 mm focal length, an 810/10 nm narrowband filter and a single photon count model (SPCM-AQRH-FC, Excelitas Technologies), respectively. The coincidence photon counts between $D_1$ and $D_2$ are determined by a coincidence circuit with a time window of 16 ns. The DMD is made up of a 768×1024 addressable digital micro-mirror array with each pixel size of 13.68 μm×13.68 μm and the relevant controlling circuit, which can modulate the spatial distribution of light intensity (i.e., photons) according to the displayed pattern. A trigger signal will be generated synchronously by the controlling circuit once a pattern is impressed on the DMD. Then the trigger signal and the coincidence photon signal are fed to a time correlated single photon count (TCSPC) module (HydraHarp 400, Picoquant) and their time sequence is recorded, so the sum of coincidence photon counts of each pattern in arbitrary interval can be integrated. The single photon counts rate of $D_1$ ($D_2$) is about $8\times10^4$/s ($3\times10^4$/s) and the coincidence photon counts rate is about 400/s, while a full-white pattern is displayed on the DMD.

According to the first order perturbation theory, the output bi-photon state function in our quantum system can be expressed as following[25]

$$|\Psi\rangle = \int d_{x_s} d_{x_i} \psi(x_s, x_i) \hat{a}_s^+(x_s) \hat{a}_i^+(x_i) |0,0\rangle \qquad (1)$$

Here, $a(x)^+$ indicates the creative operator at position presentation, the subscript *i, s* refer idler and signal, is the bi-photon probability amplitude which is often approximated as $\delta(x_i + x_s)$ and indicates the entangled photon pair posses strong spatial correlation. According to the Klyshko explanation[26], when the Gaussian thin-lens equation $1/(d_1 + d_2) + 1/d_3 = 1/f$ is obeyed, the integrated coincidence photon counts between $D_1$ and $D_2$ are given by[12, 13]

$$C \propto \sum_{n=1}^{N} |A(-x_n)|^2 \left|T_n(-\frac{x_n}{\beta})\right|^2 \qquad (2)$$

Here, $|A|^2$ is the pattern loaded on the DMD, $T^2$ is the image of object cast on DMD, $x_n$ refers the *n-th* pixel in the micro-mirror array, and $\beta=(d_1+d_2)/d_3\approx2.54$, is the magnification factor of our imaging system. Therefore, the spatial information of object and pattern displayed on DMD are non-locally contained in the coincidence photon counts, which is the basement of compressive quantum ghost imaging.

During the experiment, we choose the partial Hadarm matrix $A_{M \times N}$ (M<<N, N=64×128) as sensing matrix whose entry is 0 or 1, then alternatively imprint the pattern $A_m^+ = A_m$ (64×128, the two-dimension form of the *m-th* row within $A_{M \times N}$) and its inverted counterpart $A_m^- = 1 - A_m$ onto the DMD. The corresponding coincidence photon counts are denoted as $C_m^+$ and $C_m^-$. The frame frequency of DMD is set up to 0.1 Hz, so the maximum coincidence integrated interval of each pattern is 10 s. Let us denote $A_m^{+/-} \equiv |A_m^{+/-}|^2$ and $T_n \equiv |T_n|^2$, after total *2×M* realization of frames (M pairs of complementary frames), the equation (2) can be rewritten as

$$C_{M \times 1}^+ = A_{M \times N}^+ T_{N \times 1} \tag{3.a}$$

$$C_{M \times 1}^- = A_{M \times N}^- T_{N \times 1} \tag{3.b}$$

Here, $C_{M \times 1}^{+/-}$ is the measurement vector, and $A_{M \times N}^{+/-}$ is the so called sensing matrix. Similar with the approach applied in references[16-18], we construct the +1/-1 type sensing matrix and corresponding measurement vector as blow.

$$\Delta A_{M \times N} = A_{M \times N}^+ - A_{M \times N}^- \tag{4.a}$$

$$\Delta C_{M \times 1} = C_{M \times 1}^+ - C_{M \times 1}^- \tag{4.b}$$

$$\Delta C_{M \times 1} = \Delta A_{M \times N} T_{N \times 1} \tag{4.c}$$

Based on CS optimal algorithm, the unknown image information $T_{N \times 1}$ can be retrieved by solving the linear equation (3) and (4.c).

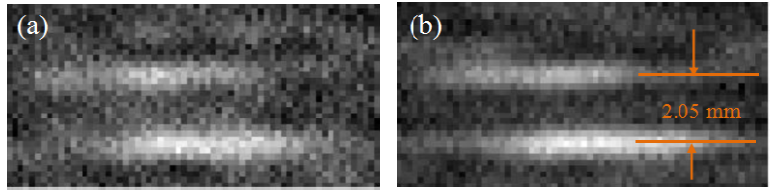

Fig.2. (a) and (b) are the reconstructed images with OMP algorithm, employing non-normalized and normalized measurement vector, respectively.

The key point of image reconstruction is the measurement signal (i.e, the fluctuation of coincidence photon counts) brought by the random patterns displayed on DMD. In the experiment, we observe that the coincidence photon counts drift systematically, which is resulting from the quantum efficiency fluctuation of SPCM induced by environmental temperature controller varying ($\pm 1.5°C$)[27]. It should be pointed out that $C_m^+$, $C_m^-$ and the sum of them $S_m$ have same drift trend in the whole sampling process. Because the sum of complementary $A_m^+$ and $A_m^-$ is a full-white pattern, the measurement independent $S_m$ is able to indicate the temperature fluctuation. We can regard $S_m$ as a reference value to normalize the measurement vector as below.

$$C_{m\_norm}^+ = \frac{C_m^+}{S_m} \cdot \overline{C} \tag{5.a}$$

$$C_{m\_norm}^- = \frac{C_m^-}{S_m} \cdot \overline{C} \tag{5.b}$$

$$\overline{C} = \frac{\sum_{m=1}^{M} S_m}{M} \qquad (5.c)$$

Fig. 2(a) and Fig. 2(b) are the reconstructed results employing non-normalized and normalized measurement vector by OMP algorithm, respectively, when M=500 and the integrated interval is 10 s. In order to speed the reconstruction process, only the central part 32×64 of each pattern is used. Thus, the final reconstructed image is 32×64 with each pixel size of 164.16 μm×109.44 μm. It is obvious that Fig.2 (b) is clearer than Fig.2 (a), which suggests that the influence of temperature fluctuation is partially offset via normalization. Therefore, the normalized measurement vector is applied in our following discussion. Meanwhile, we note that the double slit separation is about 2.05 mm (12.5 pixels), which indicates the actual magnification of our imaging system equals 2.56. It is agree well with the theoretical magnification.

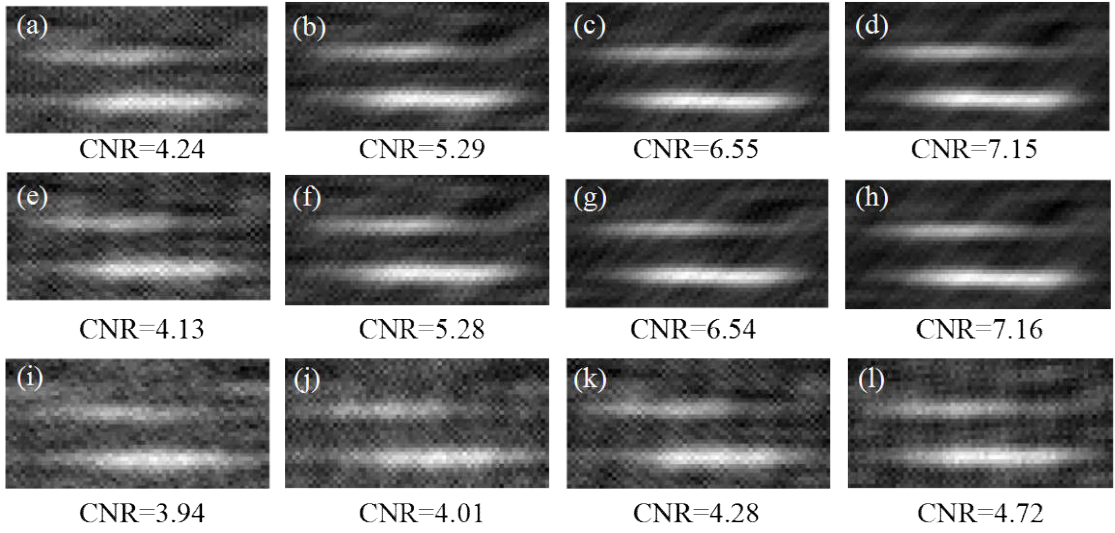

FIG.3. With OMP algorithm, (a)-(d), (e)-(h), (i)-(l) are the reconstructed images employing sensing matrix $A^+$, $A^-$ and $\Delta A$ under different measurement numbers 400, 600, 800, 1000.

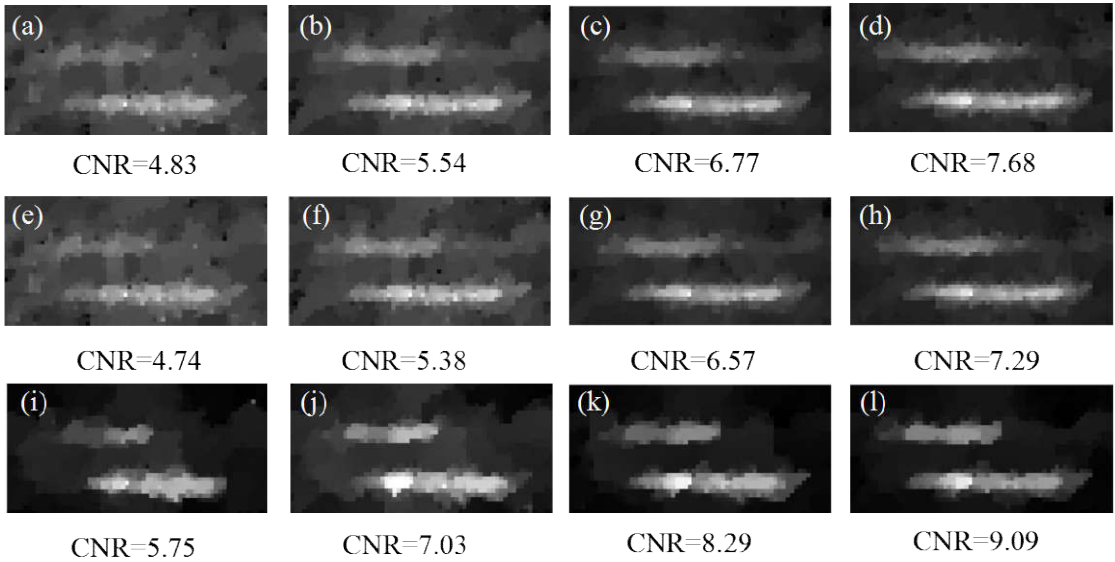

FIG.4. With TVAL3 reconstructed algorithm (a)-(d), (e)-(h), (i)-(l) are the reconstructed images employing sensing matrix $A^+$, $A^-$ and $\Delta A$ under different measurement numbers 400, 600, 800, 1000.

Based on the abovementioned three kinds of sensing matrixes (i.e., $A^+$, $A^-$ and $\Delta A$) and their corresponding normalized measurement vectors, we apply OMP and TVAL3 algorithms to retrieve the image under different measurement numbers (i.e, 400, 600, 800, and 1000) and 10 s of integrated interval. The corresponding sampling ratio of raster scanning is 19.53%, 29.29%, 39.06%, and 48.82%, respectively. The experimental results are shown in Fig. 3 and Fig. 4 respectively, and the image quality is quantitatively characterized in terms of Contrast-to-Noise ratio (CNR) [28, 29]. For an ideal image, the amount of photons falling into the double slit should be above zero, while the amount is zero in the background part. However, in the actual reconstructed image, the intensity of background part floats around zero. Therefore, we set the absolute value of the minimum intensity as a threshold $T_t$, and the intensity below and above $T_t$ is regarded as background $T_b$ and signal $T_s$ respectively. Thus, the CNR can be calculated as below.

$$CNR = \frac{aver(T_s) - aver(T_b)}{std(T_b)} \qquad (6)$$

Here, *aver* means average of $T_s$ and $T_b$, *std* is the standard deviation of $T_b$. Generally, the higher CNR, the better image quality is. The image quality degrades just as expected with the measurement numbers decreasing in both Fig. 3 and Fig. 4, no matter which sensing matrixes and reconstruction algorithms are employed. In Fig. 3, the quality of images (i)-(l) are worse than their counterparts (a)-(d) and (e)-(h) at the same numbers of frames realization. Comparing the image Fig. 3 (k) with Fig. 3(a), we notice that the constructed image quality is not significantly improved even though -1/1 sensing matrix and the same incoherent measurement numbers (i.e, 800) are applied. For the (i)-(l) in Fig. 4, they not only outperform their counterparts (a)-(d) and (e)-(h) in Fig. 4, but also have higher quality than the counterparts in Fig. 3. Hence, the RIP required by TVAL3 algorithm is stricter in comparison with OMP algorithm, and it is the type -1/1 sensing matrix that makes the optimal algorithm gain better performance. This proves that TVAL3 is more suitable for CCI technique.

Besides the measurement numbers, the shot noise (i.e., the intrinsic fluctuation of photon counts) is another major factor influencing the image quality in our single-photon level imaging scheme[30]. Once the measurement signal is overturned by the shot noise, this will cause image reconstruction to fail. Normally, the influence of shot noise can be weakened by extending the integrated interval of each pattern. As a result, it is always time-consuming and the photon utilization efficiency is decreased. To further investigate the photon utilization efficiency of CCI technology, we construct the image employing $A^+$ and $\Delta A$ by TVAL3 algorithm under the fixed measurement numbers (i.e., 500) and different photons per measurement (i.e., 800, 1600, 2400, and 3200 by setting integrated interval as 2 s, 4 s, 6 s, and 8 s), and the experimental results are shown in Fig. 5. It can be seen that image possessing more photons has higher quality when the same type sensing matrix is employed. The images (e)-(h) employing CCI are clearer than their counterparts (a)-(d) at the same amount of photons. Herein, the CCI-based imaging system has stronger robustness to the shot noise in the single photon level, which means we can obtain the same quality image using fewer photons and enhance the information-load capability of each photon by CCI technology. Note that it is possible to improve the performance (e.g. sampling time, quality of image) of our imaging system can by enhancing the brightness of entangled-photon source and optimizing the temperature controller.

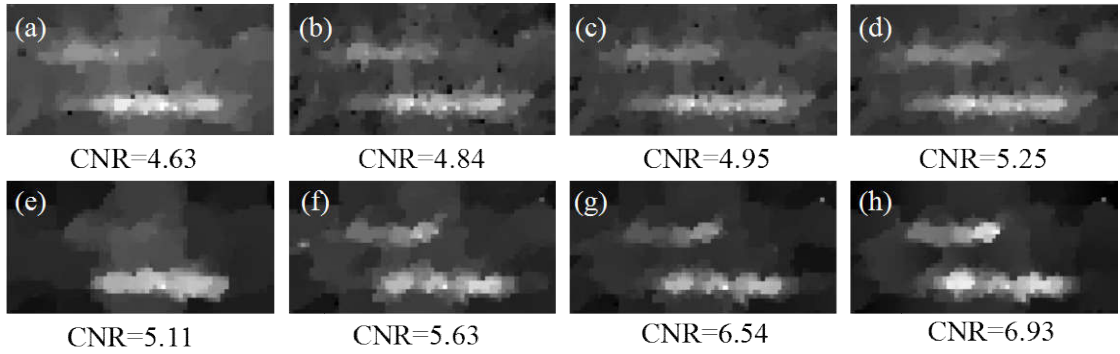

Fig.5. With TVAL3 algorithm, (a)-(d) and (e)-(h) are the reconstructed images employing $A^+$ and $\Delta A$ under 800, 1600, 2400, 3200 photons/measurement.

In conclusion, we have demonstrated a compressive normalized E-type GI system with CCI technique. The influence of photon counts drift during the measurement process is diminished by normalizing the measurement vector. The quantum ghost image reconstruction was achieved with only 19.53% sampling ratio. Comparing the quality of images constructed under different conditions (such as algorithms, sensing matrix and amount of imaging photons), it proves that CCI with TVAL3 construction algorithm can enhance the performance of system (robustness, sampling time, photon utilization efficiency). Our results reveal the great potential of CCI technique applied in quantum imaging and accelerate the pace of E-type GI toward practical application such as short range target detection and image encryption.

## Acknowledgement

This research is sponsored by the NSFC (61475191) and CAS "Light of West China" Program (XAB2015B24)